\documentclass[final,5p,times,twocolumn]{elsarticle}

\usepackage{graphicx}
\usepackage{caption}
\usepackage{latexsym}
\usepackage{amsmath,amssymb}
\usepackage{lipsum}
\usepackage{xcolor}
\usepackage[utf8]{inputenc}
\usepackage[english]{babel}
\usepackage{lineno}

\usepackage[colorlinks=True,linkcolor=DarkRed,citecolor=ForestGreen,urlcolor=MediumBlue,
	pdfstartview=FitH,bookmarks=False,pdfpagemode=UseNone
]{hyperref}
\usepackage{amsmath}

\addto\captionsenglish{}
\usepackage{bibunits} % for supplementary

\usepackage{titlesec}
\titleformat{\paragraph}[runin]
{\bfseries\scshape}{\theparagraph}{1em}{}

\makeatother

\begin{document}

%\linenumbers
\setlength\linenumbersep{0.1cm}

%\title{Identification of the $d$-wave vortex core in cuprate high-temperature superconductors}
\title{Vortex-core spectroscopy of $d$-wave cuprate high-temperature superconductors}

\author{Ivan Maggio-Aprile}
\author{Tejas Parasram Singar}
\author{Christophe Berthod}
\author{Tim Gazdić}
\author{Jens Bruér}
\author{Christoph Renner}
\affiliation{Department of Quantum Matter Physics, University of Geneva, 1211 Geneva, Switzerland}

\date{\today}

\begin{abstract}
The mechanism of high-temperature superconductivity remains one of the great challenges of contemporary physics. Here, we review efforts to image the vortex lattice in copper oxide-based high-temperature superconductors and to measure the characteristic electronic structure of the vortex core of a $d$-wave superconductor using scanning tunneling spectroscopy.
\end{abstract}

\maketitle

\section{Introduction}
The observation of superconductivity at an unprecedented high temperature in Ba$_x$La$_{5-x}$Cu$_5$O$_5$ over 35 years ago \cite{Bednorz1986}, has triggered considerable theoretical and experimental efforts to elucidate the underlying electron-pairing mechanism. To this date, high-temperature superconductivity (HTS) in copper oxide compounds remains a very active and challenging area of research. Among the many techniques used to investigate HTS, scanning tunneling microscopy (STM) and scanning tunneling spectroscopy (STS) have made important contributions, measuring the superconducting gap amplitude and symmetry, characterizing the pseudogap phase and excitation spectrum of quasiparticles, exploring atomic-scale defects \cite{fischer2007}, as well as competing (charge-ordered)  and parent (pair density wave) phases \cite{webb2019}. 

The fundamental excitations bound to magnetic vortices in type-II superconductors carry information about essential properties of the superconducting state. Their proper identification is therefore of primary interest to elucidate the mechanism driving HTS. In 1964, Caroli, de Gennes, and Matricon \cite{caroli1964} used the Bardeen-Cooper-Schrieffer (BCS) theory of superconductivity to predict that Abrikosov vortices would host a collection of localized electron states bound to their cores. The subsequent observation of these localized states \cite{hess1989}, and their response to disorder using STS \cite{renner1991}, provided a spectacular verification of the BCS theory, affirming the existence of vortex-core bound states for an $s$-wave superconductor.

Early scanning tunneling spectroscopy maps of vortex cores in HTS were neither compatible with the discrete Caroli-de Gennes-Matricon (CdGM) bound states expected for an $s$-wave superconductor \cite{caroli1964} nor with the continuum first calculated by Wang and MacDonald for a $d$-wave superconductor \cite{wang1995}. Instead of the expected zero-bias conductance peak (ZBCP) that splits with increasing distance from the core in the $d$-wave case, these experiments revealed low-energy ($E < \Delta_{\mathrm{SC}}$, where $\Delta_{\mathrm{SC}}$ is the superconducting gap) subgap states (SGSs) in YBa$_2$Cu$_3$O$_{7-\delta}$ (Y123) \cite{maggio1995} and a pseudogap like spectrum in Bi$_2$Sr$_2$CaCu$_2$O$_{8+\delta}$ (Bi2212) \cite{renner1998}. Subsequent STS mapping with improved resolution also found the presence of SGSs in the vortex cores of Bi2212 \cite{hoogenboom2000,pan2000}. The SGSs in Bi2212 were associated with a periodic short range $\approx4a_0 \times 4a_0$ modulation of the local density of states spanning the vortex-core region, where $a_0$ is the atomic lattice parameter \cite{levy2005}. Whether the SGSs reported in Y123 are also associated with a periodic charge modulation remains an open question due to the extreme difficulty of obtaining atomic resolution STM data on this material. 

\section{Vortex imaging and core spectroscopy by STS}
\begin{figure}[b]
    \centering
    \includegraphics[width=\columnwidth] {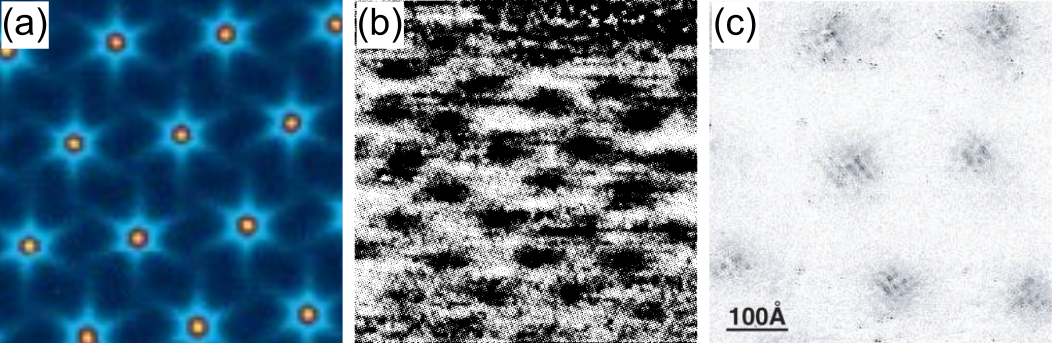}
    \caption{STS images of the vortex lattice in (a) 2H-NbSe$_2$ ($0.46 \times 0.46~\mu\mathrm{m}^2$, 0.1~Tesla; adapted from \cite{hess1990}), (b) YBa$_2$Cu$_3$O$_{7-x}$ ($0.1 \times 0.1~\mu\mathrm{m}^2$, 6~Tesla; adapted from \cite{maggio1995}), and (c) Bi$_2$Sr$_2$CaCu$_2$O$_{8+\delta}$ (difference between STS maps at 5~Tesla and at 0~Tesla; adapted from \cite{hoffman2002}).}
    \label{fig:VortexImage}
\end{figure}

High-resolution real space imaging of the Abrikosov vortex lattice by STS was first achieved on 2H-NbSe$_2$ \cite{hess1989}. These images relied on plotting the conductance as a function of position while scanning the sample surface at a tunneling setpoint corresponding to the superconducting coherence peaks at the gap edges. Much more detailed information about the complexity of the vortex-core structure was subsequently obtained by measuring full $I(V,\vec{r})$ or $dI/dV(V,\vec{r})$ tunneling spectra on a dense grid over the sample surface, and plotting the conductance at different energies below and above the superconducting gap in an XY map [Fig.~\ref{fig:VortexImage}(a)] \cite{hess1990}. STM data offer much more than high-resolution images of the vortex lattice, as they provide a unique insight into the electronic structure of the vortex core. By fitting the vortex profile in a zero-bias conductance map, it is also possible to extract fundamental superconducting quantities, such as coherence length and Fermi velocity. 

Tremendous progress has been made since the first observation of individual vortex cores by STS \cite{suderow2014}. The energy resolution in the early experiments was not sufficient to resolve the detailed CdGM electronic structure. Improved instrumentation and selected materials with a larger superconducting gap and a lower Fermi energy have since enabled to resolve the discrete CdGM states bound to the vortex core of an $s$-wave superconductor \cite{chen2018,chen2020}. However, the electronic structure of a $d$-wave vortex remained to be measured.

\section{Vortex imaging and core spectroscopy in cuprate HTS}
The observation of individual vortices in a $d$-wave HTS cuprate has remained a challenge for a long time, despite remarkable advances in scanning probe instrumentation and single crystal synthesis. This is mainly due to three factors: i) the difficulty of obtaining high-quality surfaces, which above all, exhibit homogeneous spectroscopy; ii) the absence of a salient spectral feature specific to the vortex cores in most STS data sets; and iii) a very short coherence length, which implies that vortex cores are very small and therefore easily missed by STS. These problems are exacerbated by the propensity of vortices in HTS to bind to defects, which prevents the formation of a regular lattice and necessitates disentangling the electronic structure of the defect from that of the vortex core. This last point is particularly annoying as some defects show a conductance peak at or near zero bias \cite{yazdani1999,pan2000def}, which might be mistaken for the $d$-wave ZBCP at the centre of the vortex.

\subsection{STS vortex mapping in YBa$_2$Cu$_3$O$_{7-x}$}
YBa$_2$Cu$_3$O$_{7-x}$ was the first $d$-wave cuprate high-temperature superconductor where vortices were successfully observed by STS \cite{maggio1995}. A typical spectroscopic image obtained on a fully oxygenated single crystal ($T_c=91$~K) in a magnetic field of 6~Tesla at 4.2~K is reproduced in Fig.~\ref{fig:VortexImage}(b). This was a remarkable result, since it was achieved on a thoroughly cleaned native surface (not cleaved), where atomic resolution topography was not possible. Nevertheless, tunneling spectroscopy was stable and very reproducible, with STS images revealing very small and slightly elongated vortex cores. The oval shape of the vortices is consistent with the ab-plane anisotropy in Y123, although the lack of atomic resolution has prevented identification of the local orientation of the $a$ and $b$ axes.

\begin{figure}[h]
    \centering
    \includegraphics[width=\columnwidth] {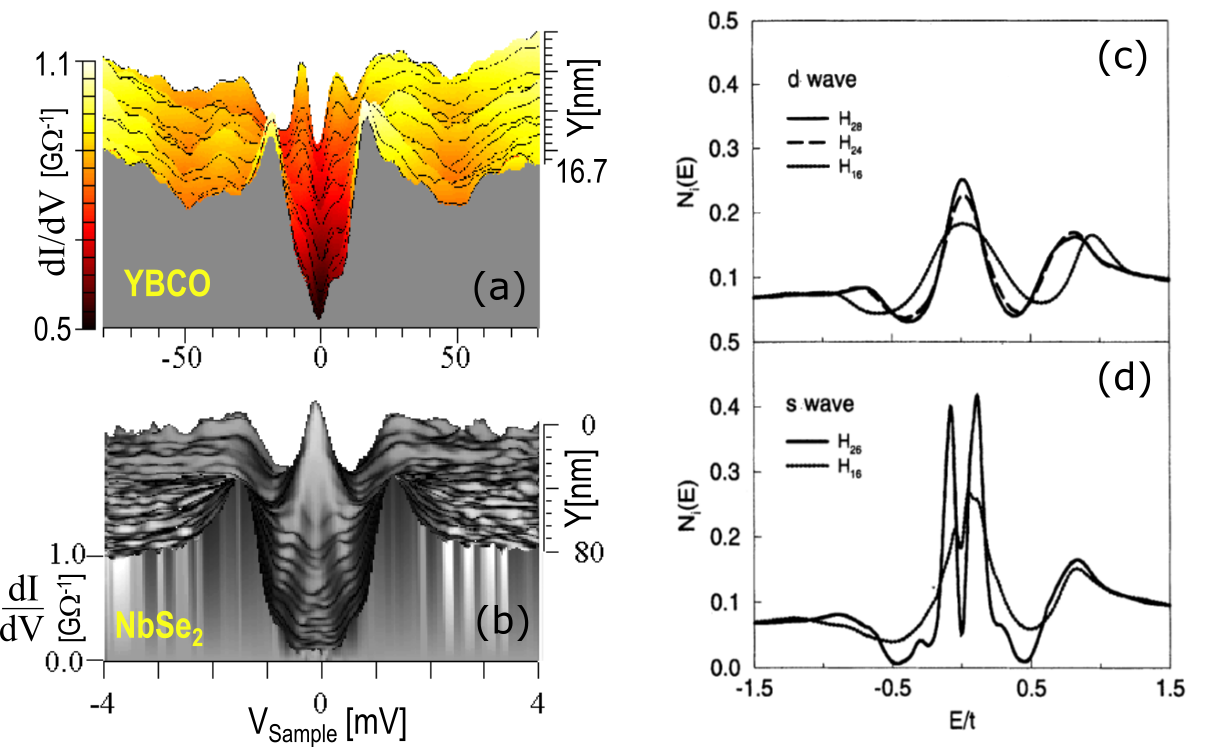}
    \caption{Tunneling conductance curves measured as a function of distance from the vortex core ($Y=0$) in (a) Y123 (adapted from \cite{maggio1995}) and (b) 2H-NbSe$_2$ (adapted from \cite{hess1989}). Corresponding model calculations for the local density of states (LDOS) at the centre of a vortex core in a (c) $d$-wave and (d) $s$-wave superconductor (adapted from \cite{wang1995}).}
    \label{fig:LowHighVortexCore}
\end{figure}

Tunneling spectroscopy at the centre ($Y=0$) of a vortex in Y123 at 4.2~K revealed two well defined low-energy conductance peaks at about $\pm 5.5$ meV \cite{maggio1995}, which do not shift in energy with increasing distance from the vortex core. As seen in Fig.~\ref{fig:LowHighVortexCore}(a), they progressively weaken while at the same time, the superconducting coherence peaks recover over a distance of the order of the coherence length $\xi$. Theory predicts a very different electronic vortex-core structure for a $d$-wave superconductor \cite{wang1995, franz1998}. Calculations find a conductance peak at zero bias at the vortex centre [Fig.~\ref{fig:LowHighVortexCore}(c)] that splits into two subgap conductance peaks, which continuously shift to higher energies with increasing distance from the core centre. This distance dependence is more like what had been observed in 2H-NbSe$_2$ [Fig.~\ref{fig:LowHighVortexCore}(b)] \cite{hess1990}. Thus, experiments seemed in contradiction with theory, where core spectra resembling those expected for a $s$-wave superconductor were measured in Y123, and spectra resembling those expected for a $d$-wave superconductor were measured in 2H-NbSe$_2$ (Fig.~\ref{fig:LowHighVortexCore}). 

The presence of a ZBCP at the centre of a vortex in 2H-NbSe$_2$ was well understood. It is a direct consequence of the very small energy splitting ($\Delta_{\mathrm{SC}}^2/E_{\mathrm{F}}$) of the lowest CdGM bound states, too small to be resolved in the experiment presented in Fig.~\ref{fig:LowHighVortexCore}(b). Subsequent studies with improved energy resolution on different $s$-wave superconductors with a larger energy spacing between adjacent CdGM bound states \cite{chen2018,chen2020}, confirmed the tunneling spectrum expected at the centre of a vortex with CdGM bound states shown in Fig.~\ref{fig:LowHighVortexCore}(d). 

\begin{figure}[b]
    \centering
    \includegraphics[width=\columnwidth] {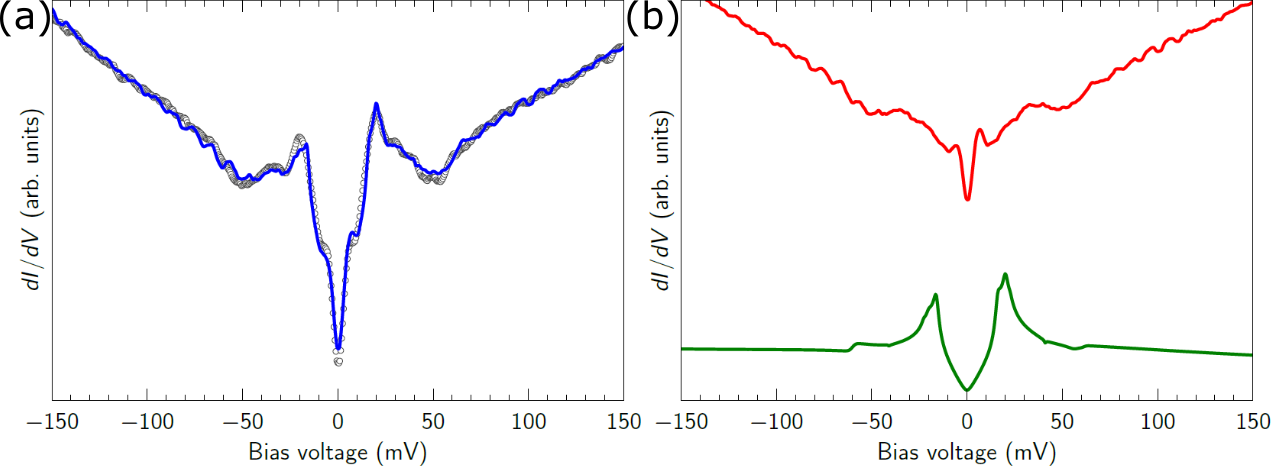}
    \caption{Phenomenological modelling of the tunneling conductance measured on Y123. (a) The solid line is the sum of the two spectra displayed in (b). It reproduces very well the experimental data measured in zero-field at 0.4~K shown as symbols. (b) Top: Vortex-core spectrum measured in a field of 6~Tesla at 0.4~K. Bottom: Calculated tunneling conductance for a $d$-wave superconductor with realistic band structure and coupling to spin excitations. (Adapted from \cite{bruer2016}).}
    \label{fig:Y-123Spectra}
\end{figure}

The puzzle of the peculiar, non-$d$-wave vortex-core spectra in Y123 was solved when a new experiment revealed that the two non-dispersing SGSs are not vortex-core states. They were found to persist throughout the sample, on and off the vortex cores, and even in the absence of any magnetic field as seen in Fig.~\ref{fig:Y-123Spectra}(a). This observation prompted Bruér \textit{et al.}\ \cite{bruer2016} to propose a phenomenological model to explain the tunneling conductance spectra measured on Y123 at 0.4~K. They were able to reproduce the zero-field tunneling conductance spectra assuming two parallel contributions [Fig.~\ref{fig:Y-123Spectra}(b)]; one from a two-dimensional band distorted by the coupling to spin excitation and subject to conventional BCS $d$-wave pairing; and one from a non-superconducting incoherent bath. 

\begin{figure}[tb]
    \centering
    \includegraphics[width=\columnwidth] {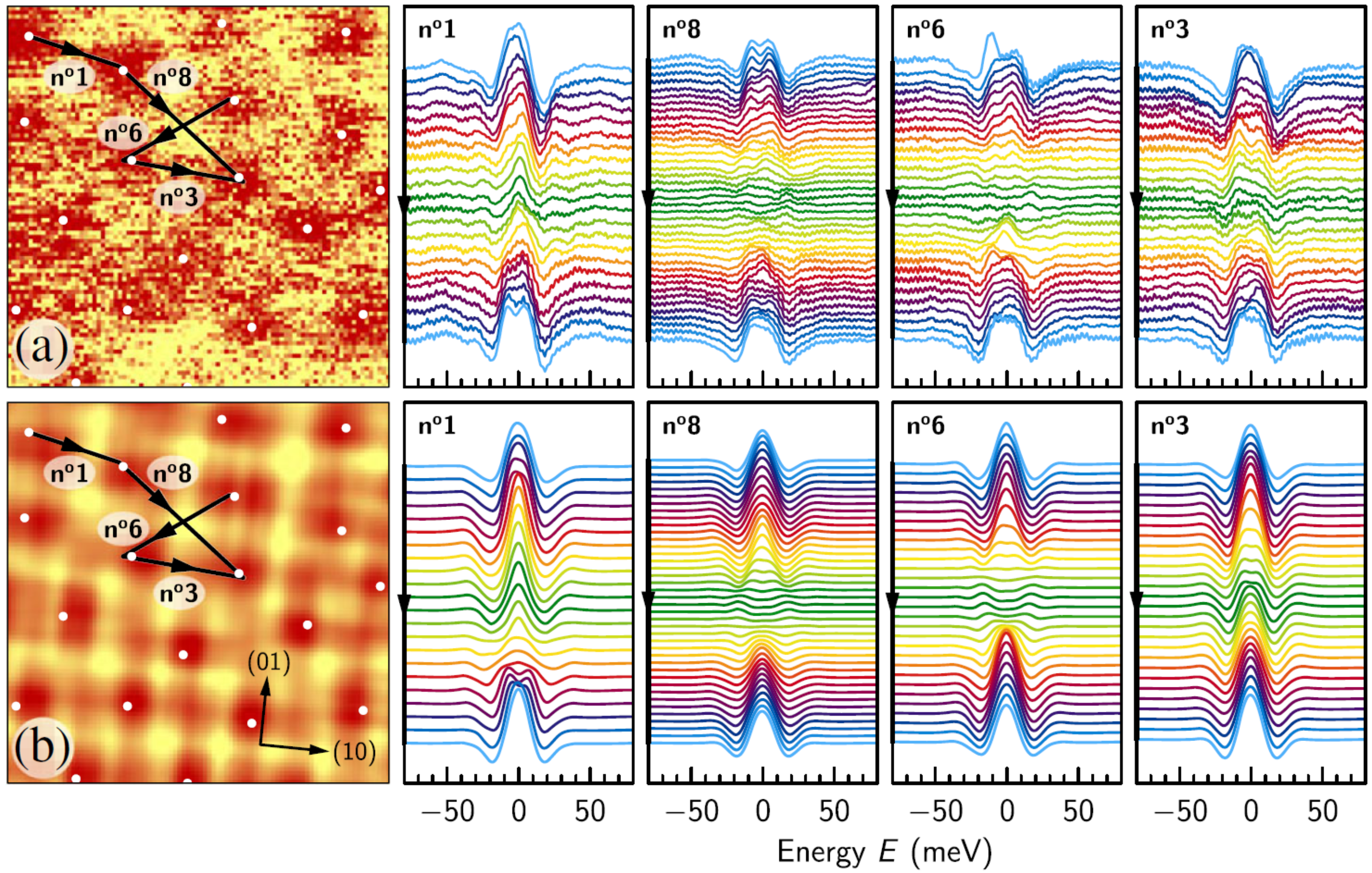}
    \caption{$90 \times 90$~nm$^2$ spectroscopic image of the vortex lattice on the (001) surface of Y123. The color scale is the ratio of the STS tunneling conductance at +5~mV and +17~mV. The four panels to the right correspond to the tunneling spectra along the indicated traces joining neighbouring vortices, divided by a tunneling spectrum away from any vortex core. (a) Experimental STS data at 6~Tesla. (b) Corresponding model calculations based on a $d$-wave superconducting gap. Note the excellent match between the data and model (see \cite{berthod2017} for details).}
    \label{fig:VortexModel_YBCO}
\end{figure}

While the phenomenological model illustrated in Fig.~\ref{fig:Y-123Spectra} does not explain the origin of the SGSs, it offers a scheme to eliminate their contribution from the vortex conductance maps, and to access the intrinsic vortex-core local density of states (LDOS). The assumption is that the non-superconducting contribution is affecting all the measured tunneling spectra, whether a magnetic field is applied or not. Hence, this unknown feature can be eliminated by subtracting a zero-field spectrum from every spectrum measured in an applied magnetic field. These experimental spectra can then be directly compared to theoretical ones obtained by subtracting the superconducting DOS from the vortex-core LDOS, both calculated for a superconductor with a $d$-wave order parameter \cite{berthod2017}. The actual calculations are somewhat more complicated, because one needs to take into account the finite image size and resolution, and the real vortex distribution at the sample surface. The latter is crucial, because each vortex is in a slightly different local environment, meaning that it is subject to different screening currents, which modify the LDOS. The correspondence between theory and data in Fig.~\ref{fig:VortexModel_YBCO} is remarkable, including minute differences between vortex cores and between traces along the [100] and [110] directions \cite{berthod2017}. These data and analysis were the first, although somewhat indirect, experimental evidences for the $d$-wave vortex-core structure. As we will see in the next section, recent STS studies in highly overdoped (OD) Bi$_2$Sr$_2$CaCu$_2$O$_{8+\delta}$ provide much more direct evidence for the electronic structure of a $d$-wave vortex core \cite{gazdic2021}.

\subsection{STS vortex mapping in Bi$_2$Sr$_2$CaCu$_2$O$_{8+\delta}$}

\begin{figure}[tb]
    \centering
    \includegraphics[width=\columnwidth] {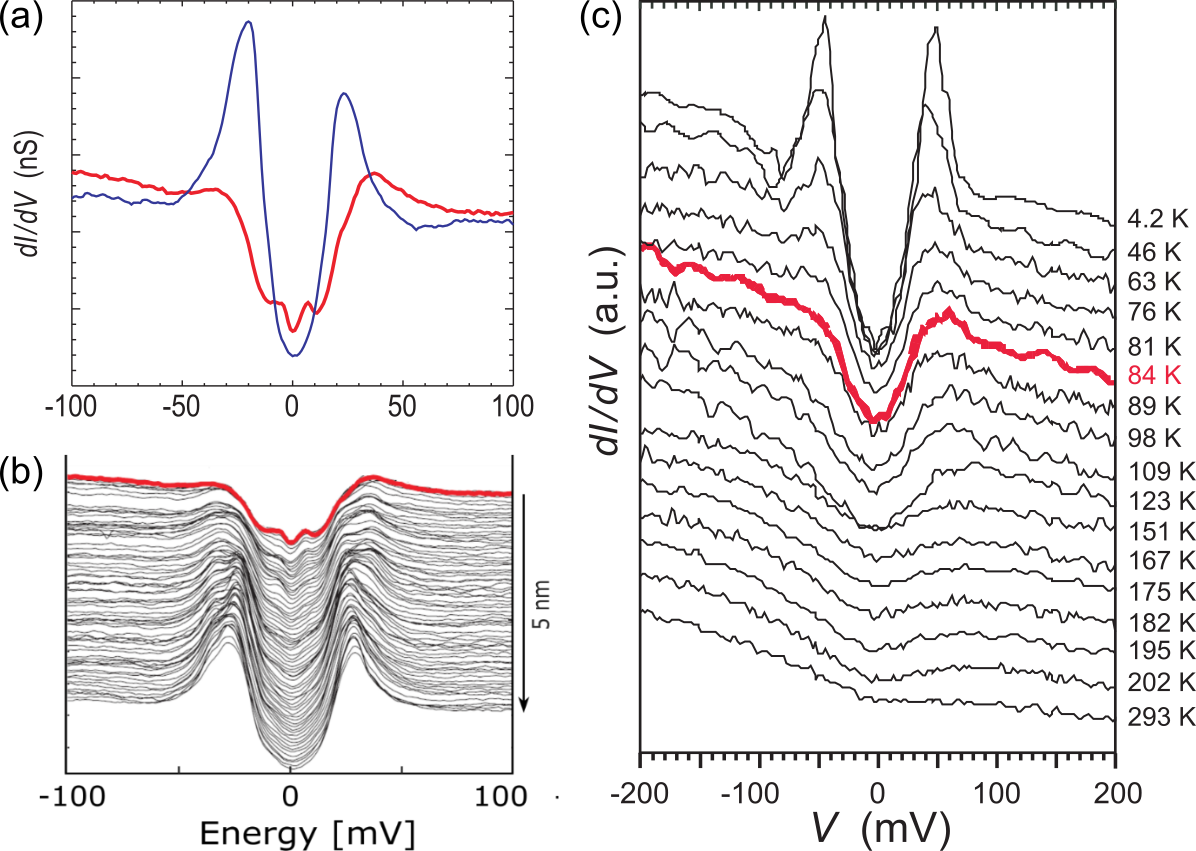}
    \caption{Tunneling spectroscopy of slightly UD Bi2212. (a) Tunneling spectra measured near the vortex centre (red) and outside the vortex cores (blue) in a field of 6~Tesla at 4.2~K (adapted from \cite{hoogenboom2000}). (b) Tunneling spectra measured as a function of increasing distance from the vortex core at 6~Tesla and 4.2~K (adapted from \cite{hoogenboom2000}). (c) Temperature-dependent tunneling conductance curves in the Meissner phase. The red spectrum corresponds to $T_c$ (adapted from \cite{renner1998}).}
    \label{fig:Bi2212UD_Core_Temp}
\end{figure}

Detecting vortices by STS in Bi$_2$Sr$_2$CaCu$_2$O$_{8+\delta}$ has been, and remains very challenging. This is surprising because, in contrast to Y123, reproducible atomic resolution STM topographic imaging and spectroscopy is routinely achieved on this compound. The first successful STS mapping of the vortex lattice in Bi2212 was reported on slightly under- and over-doped single crystals at 4.2~K and 6~Tesla \cite{renner1998}. Vortex contrast was extremely difficult to obtain, because the superconducting gap is very inhomogeneous along the sample surface --making the conductance at the coherence peak energy unsuitable to detect the vortices-- and because there is no specific spectral feature associated with the vortex. The electronic structure of the vortex cores revealed in these experiments turned out to be very different from the theoretical expectations. Instead of the ZBCP expected at their centre for a $d$-wave superconductor [Fig.~\ref{fig:LowHighVortexCore}(b)], STS revealed a LDOS very similar to the normal state DOS measured above the superconducting transition temperature $T_c$ \cite{renner1998_pg}. Follow up experiments with better energy resolution later found two SGSs below the pseudogap inside the vortex core (Fig.~\ref{fig:Bi2212UD_Core_Temp}) \cite{hoogenboom2000,pan2000}. These results have raised important questions about the pseudogap, its origin and its link with the superconducting state. The fact that the amplitudes of the pseudogap and the superconducting gap are similar and independent of temperature for different hole concentrations seems to favor a scenario where both have a common origin, consistent with an incoherent pairing state above $T_c$ \cite{emery1995, franz2001}. 

The electronic structure of the vortex cores in slightly underdoped (UD) Bi2212 became even more intriguing when Hoffman \textit{et al.}\ \cite{hoffman2002} reported a $\approx 4a_0 \times 4a_0$ periodic charge modulation oriented along the Cu-O bonds in the vortex halo region [Fig.~\ref{fig:VortexImage}(c)]. This so-called \textit{checkerboard} was initially associated with some ordered electronic phase linked to the pseudogap \cite{mcelroy2003}. It was later found that the checkerboard spanning the vortex halo in these compounds had both a low-energy dispersing, and a higher-energy (above $\Delta_{\mathrm{SC}}$) non-dispersing character. The dispersing component was identified as a vortex enhanced quasiparticle interference (QPI) in Ca$_{2–x}$Na$_x$CuO$_2$Cl$_2$ \cite{hanaguri2009}, and the non-dispersing component as a broken-spatial-symmetry state such as a CDW \cite{machida2016}. Note that in UD and nearly optimally-doped Bi2212, the dispersing checkerboard is very similar to the periodic charge modulation found throughout these samples in the Meissner phase, and away from the vortex cores in the mixed phase \cite{mcelroy2003,wise2008, machida2016}, while the non-dispersing component is similar to the one observed above $T_c$ \cite{vershinin2004}. Angle-resolved photoemission, on the other hand, failed to find any evidence for a CDW, and explained the periodic charge modulations observed at all energies in the Meissner phase in terms of QPI \cite{chatterjee2006}. 

A number of theoretical studies were conducted in order to explain why the ZBCP expected for a $d$-wave superconductor was systematically missing in the STS experiments. Since, until recently, most of the measurements were performed in UD and near optimally-doped Bi2212, the dominant suspicion was that the vortex-core signatures had in a way or the other to do with the pseudogap phase. Indeed, several theoretical explanations pointed the emergence of a competing non-superconducting static order \cite{Andersen-2000, Zhu-2001a, Maska-2003} or the resurgence of strong dynamical pseudogap-state correlations \cite{Berthod-2001b} as the phenomena that remove the low-energy states from the vortex core. Other works invoked the development of a fully gapped secondary superconducting component \cite{Himeda-1997, franz1998, Tsuchiura-2003, Fogelstrom-2011}, or extrinsic effects like the anisotropy of c-axis tunneling \cite{Wu-2000}.

\begin{figure}[tb]
    \centering
    \includegraphics[width=\columnwidth] {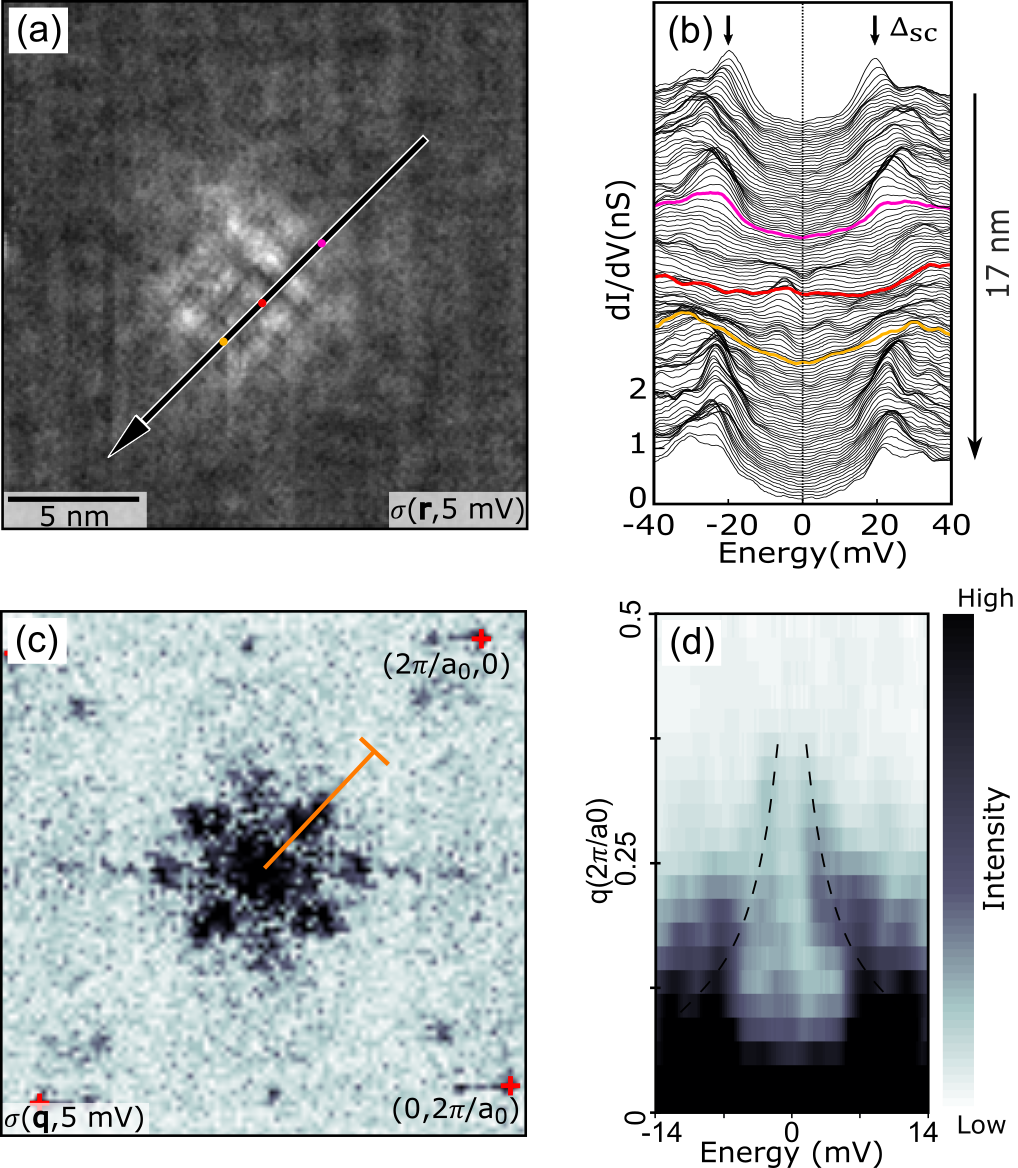}
    \caption{Electronic structure of a checkerboard vortex core in heavily-OD Bi2212 ($T_c \approx 52$~K) at 3~Tesla and 4.2~K. (a) STS conductance map at 5~mV revealing a vortex core and its checkerboard. (b) Tunneling conductance spectra measured along the arrow in panel (a). The red spectrum corresponds to the vortex centre. The pink and orange spectra delimit the core region where the SGSs develop and the superconducting coherence peaks are suppressed. The conductance scale corresponds to the lowest spectrum, the others are offset for clarity. (c) Fourier transform of the conductance map in (a). The red crosses correspond to the lattice Bragg peaks. (d) Momentum cuts along the orange line in (c) as a function of energy. The dashed line highlights the dispersion of the checkerboard $q$-vector. (adapted from \cite{gazdic2021}).}
    \label{fig:Bi2212OD_Core_CB}
\end{figure}

\begin{figure}[tb]
    \centering
    \includegraphics[width=\columnwidth] {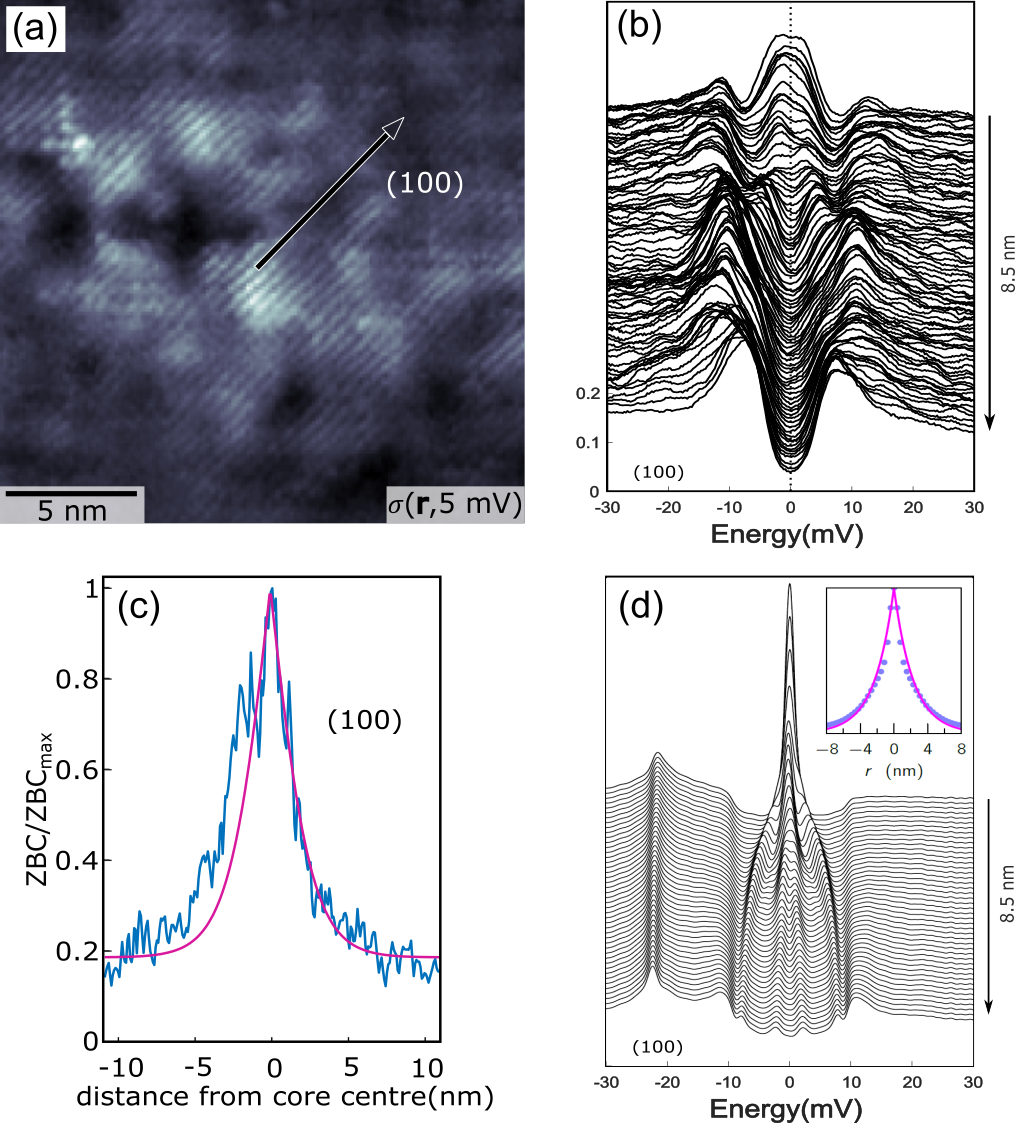}
    \caption{Electronic structure of a pristine $d$-wave vortex core in heavily-OD Bi2212 ($T_c \approx 52$~K) at 0.16~Tesla and 4.2~K. (a) STS conductance map at 5~mV revealing a vortex core. There is no checkerboard, the background fluctuations correspond to inhomogeneities. (b) Tunneling conductance spectra measured along the arrow in panel (a) revealing the $d$-wave core structure predicted by Wang and MacDonald \cite{wang1995}. (c) Normalized zero-bias tunneling conductance across the vortex centre along the (100) direction (blue) and the corresponding calculated profile $f(r)$ (pink). (d) LDOS along the (100) direction for a self-consistent isolated vortex in a tight-binding model \cite{norman1995} with doping $p=0.23$ and $d$-wave gap $\Delta=10$~meV. The energy resolution is 1~meV. The inset shows the zero-energy LDOS as a function of distance $r$ from the vortex-core and a fit $\sim \exp(-r/2.3~\mathrm{nm})$.}
    \label{fig:Bi2212OD_Core_Model}
\end{figure}

Until 2021, all STM/STS studies of Bi2212 revealed this $\approx 4a_0 \times 4a_0$ periodic charge modulation in the vortex halo. The modulation amplitude was strongest in conductance maps measured close to the SGSs energy, suggesting a link between checkerboard and SGSs. All these results were obtained on UD to slightly overdoped (OD) Bi2212 crystals, which have a pseudogap above $T_c$, further suggesting a link between the checkerboard and the pseudogap as mentioned above. A logical follow-up was to focus on heavily-OD Bi2212 ($T_c \approx 52$~K), as it should behave more like a Fermi liquid, with no pseudogap or CDW reconstruction above a critical doping $p_c\approx0.19$ \cite{norman2010}. However, much to their surprise, Gazdi\'{c} \textit{et al.}\ \cite{gazdic2021} observed the same vortex-core signature at 3~Tesla and 4.2~K (Fig.~\ref{fig:Bi2212OD_Core_CB}), with dispersing SGSs associated with a $\approx 4a_0 \times 4a_0$ periodic charge modulation. The high-field data on heavily-OD Bi2212 by Gazdi\'{c} \textit{et al.}\ \cite{gazdic2021} remove any possible link between the pseudogap and SGSs or between the pseudogap and checkerboard. One notable difference to the earlier experiments was the absence of any checkerboard in zero magnetic field (except near some defects), unlike the reentrant CDW observed by Peng \textit{et al.}\ \cite{peng2018} in OD Bi2201 outside the pseudogap regime.

The pristine $d$-wave vortex-core structure was ultimately detected by Gazdi\'{c} \textit{et al.}\ \cite{gazdic2021} in heavily-overdoped Bi2212 single crystals at 4.2~K and 0.16~Tesla (Fig.~\ref{fig:Bi2212OD_Core_Model}), an unprecedentedly low magnetic field for any HTS study by STS to date. The motivation for such a low field was to reduce the potential influence of neighbouring flux lines and come closer to the isolated vortex case considered in calculations. The challenge was to detect only few tiny vortices ($\xi \approx 2$~nm) in a large field of view (vortex spacing $\approx 120$~nm). The vortex core is identified at the centre of the 5~mV conductance map in Fig.~\ref{fig:Bi2212OD_Core_Model}(a). There is clearly no $\approx 4a_0 \times 4a_0$ charge modulation and the tunneling conductance spectrum at the centre of the vortex shows a clear peak structure at zero bias as expected for a $d$-wave superconductor [Fig.~\ref{fig:Bi2212OD_Core_Model}(b)]. Careful checks were performed to ensure the tunneling conductance features were due to the presence of a vortex, and not the spectral signature of a defect which can also feature a ZBCP. 

The evolution of the tunneling conductance with increasing distance from the vortex centre in Fig.~\ref{fig:Bi2212OD_Core_Model}(b) closely follows the theoretical expectations for a $d$-wave vortex core in Fig.~\ref{fig:Bi2212OD_Core_Model}(d), with an increasingly large splitting of the peak at the Fermi level. Interestingly, the evolution of the tunneling conductance is slightly different along the nodal (110) and antinodal (100) directions, reflecting a small anisotropy in the Fermi velocity. We emphasize that this four-fold symmetric shape of the vortex-core is a consequence of the Fermi surface topology; it is not due to the $d$-wave symmetry of the superconducting gap (see supplemental material of reference \cite{berthod2017}). The decay of the normalized zero-bias conductance $f(r)$ allows one to extract the coherence length from $f(r)=1-(1-\sigma_0)\tanh(r/\xi)$ [pink line in Fig.~\ref{fig:Bi2212OD_Core_Model}(c)], where $\sigma_0$ is the residual conductance far from the centre, $r$ is the distance from the centre and $\xi$ is the coherence length. The result is $\xi \approx2.7$~nm, implying an average perpendicular Fermi velocity of $v_F \approx 167$~km/s consistent with band structure calculations for heavily-overdoped Bi2212. The reduced gap ratio $2\Delta_{\mathrm{SC}}/k_{\mathrm{B}}T_c$ amounts to 4.5, close to 4.3, the BCS value for a weakly-coupled $d$-wave superconductor. 

\section{Discussion and outlook}

The search for $d$-wave vortices by scanning tunneling spectroscopy in high-temperature superconductors has come a long way since the first images reported in 1995 \cite{maggio1995}. The most recent work by Gazdi\'{c} \textit{et al.}\ \cite{gazdic2021} provides a number of answers to outstanding questions while raising new ones. Finding the expected $d$-wave core structure removes some of the unusual features previously attributed to $d$-wave vortex cores that have challenged new theories of the HTS ground state. The physical parameters extracted from the vortex core displayed in Fig.~\ref{fig:Bi2212OD_Core_Model} are perfectly consistent, and place heavily-OD Bi2212 in the BCS mean-field regime. Gazdi\'{c} \textit{et al.}\ \cite{gazdic2021} further invalidate any possible link between the dispersing checkerboard order and the pseudogap, since they observe it well into the overdoped region of the Bi2212 doping phase diagram, where no pseudogap has been measured. This result favors a QPI scenario \cite{chatterjee2006} over charge order to explain the checkerboard at all energies.

Outstanding questions include the nature of the SGSs and checkerboard. While the origin of the SGSs remains unclear, there are several well established experimental facts. SGSs have been reported in Y123 \cite{maggio1995, shibata2003, shibata2010} and in Bi2212 \cite{hoogenboom2000,pan2000,matsuba2007,yoshizawa2013, gazdic2021}. Hoogenboom \textit{et al.}\ \cite{hoogenboom2001} observed a linear correlation between the SGS energy and the superconducting gap in Bi2212 at various doping levels and in optimally-doped Y123, noticing this would make it challenging to consider them as discrete CdGM bound states, which scale as $\Delta_{\mathrm{SC}}^2$. 

\begin{figure}[tb]
    \centering
    \includegraphics[width=\columnwidth] {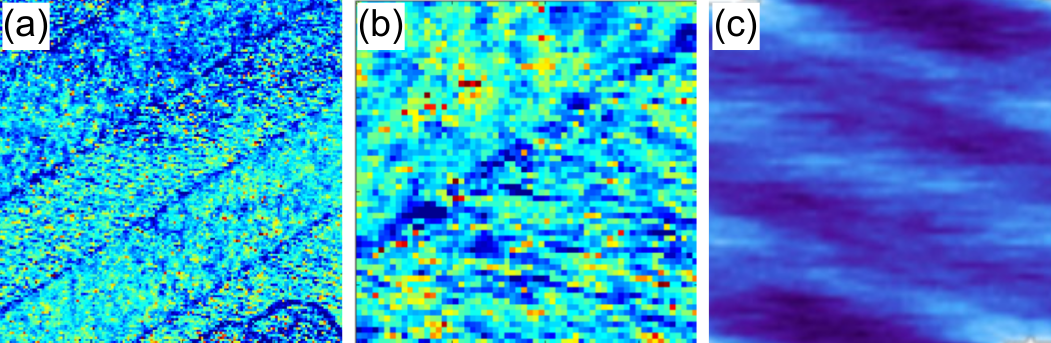}
    \caption{STS images of Y123 acquired at the SGS energy (6~mV) at 0.4~K. (a) $750 \times 750$~nm$^2$ image revealing 4 twin boundaries and alternating charge textures between them. (b) $180 \times 180$~nm$^2$ image revealing stripe like modulations aligned with the Cu-O bonds and rotated by 90\textdegree~across the twin boundary. (c) $16 \times 16$~nm$^2$ auto-correlation image of the bottom right region in panel (b) highlights a charge stripe modulation with a period of about 5~nm on top of a weaker one of about $1$~nm, both at 45\textdegree~to the TB (adapted from \cite{bruerphd}).}
    \label{fig:stripes_ybco}
\end{figure}

In Bi2212, SGSs always appear alongside the periodic $\approx$~$4a_0 \times 4a_0$ charge modulation in the vortex halo. In heavily-OD Bi2212, whenever the ZBCP is observed, there is neither any SGSs nor any checkerboard \cite{gazdic2021}. The link between the SGSs and checkerboard in Bi2212 was highlighted several years ago by Levy \textit{et al.}\ \cite{levy2005}, who noted that the checkerboard's modulation amplitude was maximal in conductance maps taken at the energy of the SGSs. This observation remains true in heavily-OD Bi2212 \cite{gazdic2021}. Some periodic charge modulations at the energy of the 6~mV SGSs in Fig.~\ref{fig:Y-123Spectra} have also been detected in optimally-doped Y123 in zero magnetic field at 0.4~Kelvin (Fig.~\ref{fig:stripes_ybco}) \cite{bruerphd}. Although the spatial resolution is far from optimal, the $750 \times 750$~nm$^2$ conductance map in Fig.~\ref{fig:stripes_ybco}(a) reveals 4 twin boundaries (TBs) along the (110) or (1\=10) crystallographic directions running diagonally through the image. Between them, one can distinguish a charge texture at 45\textdegree~to the TB. This one-dimensional texture rotating by 90\textdegree~across a TB is better seen in the $180 \times 180$~nm$^2$ image in Fig.~\ref{fig:stripes_ybco}(b). Taking an auto-correlation of the bottom right-hand region of panel (b) emphasizes stripes oriented at 45\textdegree~from the twin boundaries in Fig.~\ref{fig:stripes_ybco}(c), i.e. running along the Cu-O bonds of the CuO$_2$ planes. Whether these charge modulations are related to the ones reported from other experimental techniques \cite{wu2011,ghiringhelli2012,chang2012} remains an open question.

Another outstanding puzzle is why the $d$-wave vortex-core structure could so far only be observed in heavily overdoped Bi2212 at very low magnetic fields (Fig.~\ref{fig:Bi2212OD_Core_Model}), and why is the ZBCP replaced with SGSs and a checkerboard when increasing the magnetic field (Fig.~\ref{fig:Bi2212OD_Core_CB}). There is to the best of our knowledge no model calculation that predicts such a drastic change of the vortex-core structure for moderate field changes. It is possible that the modification is not due to the magnetic field, but to the local doping level --these samples are inhomogeneous, and it is not impossible that the STM tip is sensing regions only few nanometers apart with quite different doping levels on the same sample surface. Indeed, a recent study by Datta \textit{et al.}\ \cite{datta2023} discusses the development of some “Mottness” in the vortex-core region of weakly-doped compounds, which triggers not only the vanishing of the ZBCP, but also the emergence of low-energy core states \cite{datta2023}. Liu \textit{et al.}\ \cite{liu2023} similarly report the vanishing of the vortex-core ZBCP in underdoped compounds. These authors further claim that the SGSs become strongly enhanced at large fields in the presence of a pair density wave (PDW). Edkins \textit{et al.}\ \cite{edkins2019} claim to have identified a high-field-induced PDW state in the vortex halo of slightly UD Bi2212. All these studies rely on an electronic nematic or charge modulated background (CDW or PDW), which is assumed to form in the parent state of the pseudogap regime at high temperature ($T > T_c$), with superconductivity considered as a competing order emerging at $T \leq T_c$  \cite{agterberg2015, dai2018}. These results do not explain why SGSs and conductance modulations are also present in the vortex halo of heavily-OD Bi2212, which do not have any pseudogap, and how the magnetic field strength might influence these features. Despite looking carefully for PDW modulations described in these model calculations, we have so far not been able to detect any in our experiments, and this fascinating topic is the subject of continued investigations.   

\section{Acknowledgements}
K.\ A.\ M\"{u}ller and his team have given the community a fantastic system to explore with scanning probes. CR and IMA would like to thank all the collaborators who have made it possible to build and operate a number of scanning probe instruments in Geneva, all the crystal producers who have synthesized exceptional samples, and all the scientists who have contributed to this stimulating research over the years. CR, IMA and CB would like to dedicate this article to late professor \O{}ystein Fischer, who initiated the research on HTS and scanning probes in Geneva. The exploration of vortices in high-temperature superconductors at the University of Geneva has been supported by the Swiss National Science Foundation through various funding programs (research grants, NCCR MaNEP and R'Equip) and by the DQMP (formerly DPMC).

\bibliography{VortexHTS_Maggio-Aprile.bib}

\end{document}